\documentclass[12pt,a4paper]{iopart}
 \pdfoutput=1
\usepackage{iopams}  
\usepackage{graphicx}
\usepackage[breaklinks=true,colorlinks=true,linkcolor=blue,urlcolor=blue,citecolor=blue]{hyperref}
\usepackage{subcaption}
\usepackage{bbm}
\usepackage{color}
\expandafter\let\csname equation*\endcsname\relax
\expandafter\let\csname endequation*\endcsname\relax
\usepackage{amsmath}
\usepackage{amsfonts}%

\begin{document}
\title{Characterizing non-deterministic noiseless linear 
amplifiers at the quantum limit}
\author{Hamza Adnane}
\address{Laboratoire de Physique Th\'eorique, Facult\'e des Sciences Exactes,\\
	Universit\'e de Bejaia, 06000 Bejaia, Algeria}
\author{Francesco Albarelli}
\address{Department of Physics, University of Warwick, Coventry 
CV4 7AL, United Kingdom}
\author{Abdelhakim Gharbi}
\address{Laboratoire de Physique Th\'eorique, Facult\'e des Sciences Exactes,\\
	Universit\'e de Bejaia, 06000 Bejaia, Algeria}
\author{Matteo G. A. Paris}
\address{Quantum Technology Lab, Dipartimento di Fisica 
	{\em Aldo Pontremoli}, Universit\`a degli Studi di Milano, I-20133 
	Milano, Italy}
\ead{matteo.paris@fisica.unimi.it}
\begin{abstract}
We address the characterisation of the gain parameter of a 
non-deterministic noiseless linear amplifier (NLA) and compare 
the performances of different estimation strategies using tools
from quantum estimation theory. At first, we show that, contrary 
to naive expectations, post-selecting only the amplified 
states does not offer the most accurate estimate. We then
focus on minimal implementations of a NLA, i.e. those obtained 
by coupling the input state to a two-level system, and show
that the maximal amount of information about the gain of 
the NLA is obtained by measuring the whole composite system.
The quantum Fisher information (QFI) of this best-case scenario 
is analysed in some details, and compared to the QFI of 
the post-selected states, both for successful and unsuccessful 
amplification. Eventually, we show that full extraction of
the available information is achieved when the non-deterministic 
process is implemented by a L\"uders instrument.
We also analyse the precision attainable by probing NLAs by 
single-mode pure states and measuring the field or the number 
of quanta, and discuss in some details the specific cases 
of squeezed vacuum and coherent states.
\end{abstract}
%
%
\section{Introduction}
Deterministic, phase-insensitive, \emph{quantum} linear amplifiers 
that amplify the whole set of quadratures of a bosonic field 
unavoidably introduce additional noise~\cite{Caves} of purely 
quantum origin. In spite of this general feature, several {\em 
noiseless} amplification schemes have been proposed, both 
theoretically and experimentally~\cite{Ralph,xiang,Combes,Ferreyrol,Ferreyrol2,Usuga}.
In these setups, the ideal desired action on coherent states, i.e.
$\left\vert \alpha \right\rangle \mapsto \left\vert g 
\alpha \right\rangle$ with $g>1$, is obtained without any 
additional noise, and maintaining consistency with the basic
postulates of quantum mechanics. These {\em noiseless linear amplifier} (NLA) 
are made possible by their probabilistic nature~\cite{Ralph}, i.e.
amplification is not deterministic and it is achieved only in a 
fraction of the experimental events. In other words, NLAs are 
able to attain arbitrary high fidelities with the target amplified state, 
at the cost of obtaining a successful amplification only with a 
(usually small) probability $p_{s}$. Notice that this kind of 
devices not only do not add additional noise, but actually also 
avoid the amplification of the input noise \cite{Combes,Combes2016}.
\par
An ideal NLA, achieving perfect amplification for any input state, 
would succeed with probability zero~\cite{Menzies2009}. Therefore, 
any realistic scheme implements an approximate NLA that works with 
high fidelity for a certain class of relevant states. This approach 
introduces a trade-off between the probability of success and the 
fidelity of amplification. Pandey et al.~\cite{Combes} have identified 
the explicit form of the quantum operation that is optimal w.r.t. this 
trade-off. Building on this result, McMahon et al.~\cite{McMahon} 
found a measurement model, consisting of a unitary interaction and 
projective measurements, which implements the optimal probabilistic 
operation. In this description, the NLA is fully characterized by 
two parameters: the {\em gain} $g$ and the {\em threshold} 
$p$.

The use of NLA has been suggested in a broad range of possible 
applications to improve quantum communication protocols~\cite{Fiurasek,Ralph2,Blandino}. In this framework, a precise 
characterization of its gain would be a relevant tool to
take full advantage of this device, and a 
questions arises on whether feasible detection schemes are
available with current technology.

As a matter of fact, the gain and threshold parameters of a 
NLA do not correspond to quantum observables and, in turn, one 
has to resort to statistical estimation to infer their value.
Optimisation over the choice of a probe state, of a detection
scheme and a suitable and data processing may be performed 
in the framework of quantum estimation theory, which provides 
the ultimate bound on precision, i.e. the quantum Cramér-Rao 
bound in terms of the quantum Fisher information~\cite{Holevo2011b,helstrom1976quantum,Giovannetti,Matteo}.

NLAs are probabilistic devices and their 
characterisation is closely related to the field of 
probabilistic quantum metrology, which received much
attention in recent years~\cite{Fiurasek2,Gendra,Gendra2,Marek,Calsamiglia,Ferrie,Zhang}.
The idea behind probabilistic quantum metrology is to first 
deterministically encode a parameter onto the probe state, 
and then apply a selective measurement. By weakly measuring 
the probe state and then discarding part of the output states 
depending on the outcomes, it possible to concentrate information 
on the parameter. However, from the point of view of quantum estimation theory these protocols cannot improve the precision of the estimation in the limit of many trials~\cite{Tanaka,Combes2014}.

In this paper, we present a detailed study of the estimation of
the gain parameter of a probabilistic noiseless amplifier, focussing
attention on the NLA measurement model proposed in~\cite{McMahon}.
In particular, we will consider the NLA as a device given to an 
experimentalist who needs to calibrate it, but who cannot act 
on the building blocks. At variance with the probabilistic 
metrological protocols we have mentioned above, the non-deterministic 
nature of the NLA makes this estimation scheme \emph{intrinsically} 
probabilistic. This feature makes it necessary to consider different 
figures of merit to quantify the information obtained by measurement.
The most informative strategy is to consider both the information 
contained in the classical statistics given by the heralding process 
(i.e. the POVM implemented by the quantum instrument) and the 
information encoded in the conditional states. 
We also show that this strategy is optimal; having access to the global 
pure state after the interaction with the global unitary used to 
implement the selective evolution does not give any more information on the parameter. On the other hand, we will also consider the unconditional state, 
as well as the information encoded only in a successfully amplified state,
and compare this scheme with the optimal one.

As we will see, our analysis is general enough to assess the performances 
of any single-mode pure state used as a probe for the NLA gain. In addition,
in order to offer some quantitative assessment, we evaluate explicitly
the bounds to precision for squeezed vacuum and coherent states.

The paper is organized as follows.
In Section~\ref{s:QET}, we review the main results on single-parameter 
quantum estimation theory. Section~\ref{s:NLA} is devoted to a brief 
review of the model of non-deterministic NLA proposed by Pandey et al.
We focus on the action on generic one mode bosonic pure states and we 
also discuss its unitary dilation (the measurement model 
given by MacMahon et al.). In Section~\ref{s:CRB}, we present three 
different metrological strategies to infer the value of the gain, 
assuming a known threshold. We assess their performances in terms 
of their respective Cramér-Rao bounds and we present our main 
results, a comparison between these strategies considering two 
classes of single mode bosonic pure states: squeezed vacuum and 
coherent states. In Section~\ref{s:FI} we analyze feasible 
measurement schemes saturating the quantum bounds on the precision 
of the NLA gain. Finally, Section~\ref{s:out} closes the paper with 
some concluding remarks.
\section{Quantum estimation theory}
\label{s:QET}
Various crucial quantities for the characterization of quantum systems, for instance entanglement~\cite{Genoni2008b} or the loss parameter of a quantum channel~\cite{Monras2007,Rossi}, are non-linear functions of the density matrix and thus cannot correspond to quantum observables.
In order to have access to these quantities, one should resort to indirect measurements and set the problem in the context of quantum parameter estimation theory~\cite{helstrom1976quantum,Holevo2011b,Matteo,Braunstein}.
In the classical case, the typical 
estimation procedure is set as follows.
We intend to estimate the value of a parameter $g$, from a set of measurement outcomes $\left\{  x_{1},x_{2}....,x_{m}\right\}$, which represent a sample from a parameter-dependent probability distribution $p(x|g)$, also called the statistical model.
The collected data is processed (classically) to build an estimator 
$\tilde{e}_{g}=\tilde{e}_{g}(\left\{  x_{1},x_{2}....,x_{m}\right\} )
$, i.e. a function from the set of measurement outcomes to the set 
of possible values of the unknown parameter.
Being a function of random variables, the estimator itself is a random variable and the precision of the estimation can be quantified by its variance $\sigma^{2}(\tilde{e}_{g}) = \mathbbm{E} \left[ \left(\tilde{e}_{g} -  \mathbbm{E} \left[  \tilde{e}_{g} \right] \right)^2 \right]$.
This is a good figure of merit if we assume that the estimator is \emph{unbiased}, i.e. the average value is equal to the \emph{true} value of the parameter $\mathbbm{E} \left[  \tilde{e}_{g} \right] = g$.
An important classical result is that the variance of any unbiased estimator has a lower bound independent on the particular estimator; this is the so-called Cram\'{e}r-Rao bound (CRB)~\cite{lehmann_theory_1998}
\begin{equation}
\sigma^{2}(\tilde{e}_{g})\geq\frac{1}{M\mathcal{F}(g)}, \label{1}%
\end{equation}
where $M$ is the number of measurements performed on the system and $\mathcal{F}(g)$ the Fisher information (FI) defined as%
\begin{equation}
\mathcal{F}(g)=\sum_{x}p(x\vert g)\left[  \partial_{g}\ln p(x\vert g)\right]  ^{2}.
\label{2}%
\end{equation}
\par
In quantum theory, the conditional probability is given by the Born rule  $p(x\vert g)=\mathrm{Tr}[\rho_{g} \hat{\Pi}_{x}]$, where the set of density matrices $\rho_{g}$ now constitutes a \emph{quantum} statistical model parametrized by $g$, while $\Pi_{x}$ represents an element of the positive operator-valued measure (POVM) describing the measurement, satisfying $\sum_x \hat{\Pi}_x = \mathbbm{1}$.
Maximizing the FI over all possible POVMs we obtain the ultimate bound on the accuracy of any unbiased estimator, the quantum Cram\'{e}r-Rao bound (QCRB)~\cite{Braunstein,helstrom1976quantum}
\begin{equation}
\sigma^{2}(\hat{e}_{g})\geq\frac{1}{M\mathcal{F}(g)}\geq\frac{1}{M \mathcal{Q}(g)},
\label{3}%
\end{equation}
where $\mathcal{Q}(\rho_g)=\mathrm{Tr}\left[  \rho_{g} \hat{L}_{g}^{2}\right] $ is the so-called quantum Fisher information (QFI); $\hat{L}_{g}$ is the symmetric logarithmic derivative, an Hermitian operator defined implicitly through the equation $\partial_{g}\rho_{g}=\frac{1}{2}(\hat{L}_{g}\rho_{g}+\rho_{g} \hat{L}_{g})$.
The QFI depends only on the quantum state $\rho_{g}$, thereby setting the ultimate limit on accuracy of any estimation strategy for $g$.

For a generic mixed state a general formula of the QFI is the following~\cite{Matteo}%
\begin{equation}
\mathcal{Q}(\rho_g)=2\sum_{i,j}\frac{\left\vert \left\langle \psi_{i}\vert\partial_{g}\rho
	_{g}\vert\psi_{j}\right\rangle \right\vert ^{2}}{v_{i}+v_{j}}, \label{4}%
\end{equation}
where we used the eigendecomposition of the density matrix $\rho_{g}=\sum v_{n}\left\vert
\psi_{n}\right\rangle \left\langle \psi_{n}\right\vert $; the sum includes all $j$ and $i$ satisfying $v_{i}+v_{j}\neq0$.
For a pure state $\rho_{g}=|\psi_g \rangle \langle \psi_g |$ Eq.~(\ref{4}) reduces to%
\begin{equation}
\mathcal{Q}(| \psi_g \rangle)=4\left[  \left\langle \partial_{g}\psi_{g}\vert\partial_{g}\psi
_{g}\right\rangle - \left| \left\langle \partial_{g}\psi_{g}\vert\psi_{g}\right\rangle \right|^{2} \right] .  \label{eq:QFIpure}%
\end{equation}

\section{Description of the non-deterministic NLA}
\label{s:NLA}
Performing a phase insensitive amplification on a generic quantum state is well known to inevitably insert additional noise, and thus poses limits to quantum communication and metrology protocols.
Nonetheless, it has been shown that non-deterministic noiseless linear amplifiers can circumvent these limitations~\cite{Ralph,xiang}.
Through this manuscript, we follow the theoretical model developed in Refs.~\cite{Combes,McMahon} and we explore different strategies to calibrate such an amplifier, i.e. to precisely estimate the gain, assuming a known value of the integer $p$ setting the truncation order (which we dub the threshold).

The action of the optimal probabilistic NLA is described by two Hermitian and commuting Kraus operators~\cite{McMahon}:
\begin{eqnarray}
\hat{E}_{s}^{p}=g^{-p}\sum\limits_{n=0}^{p}g^{n}\left\vert n\right\rangle
\left\langle n\right\vert +\sum\limits_{n=p+1}^{\infty}\left\vert
n\right\rangle \left\langle n\right\vert , \label{6} \\
\hat{E}_{f}^{p}=\sqrt{1-\hat{\Pi}_{S}^{p}}=\sum\limits_{n=0}^{p}%
\sqrt{1-g^{2(n-p)}}\left\vert n\right\rangle \left\langle n\right\vert ,
\label{7}
\end{eqnarray}
where $s$ and $f$ denote respectively success and failure and the basis $|n \rangle$ is the usual Fock basis for a bosonic mode.
These Kraus operators correspond to a POVM with two outcomes, i.e. two positive operators $\hat{\Pi}_{s}^{p} = \hat{E}_{s}^{p}{}^\dag \hat{E}_{s}^{p}$ and $\hat{\Pi}_{f}^{p}=\hat{E}_{f}^{p}{}^\dag \hat{E}_{f}^{p}$ which constitute a resolution of the identity.
Here, $g$ is the gain of the amplifier and $p$ is the threshold.
The probabilities that an outcome occurs when a measurement is performed on a state $\rho$ and the corresponding post-measurement states read
\begin{equation}
p_{i}=\mathrm{Tr}\left[  \rho\hat{\Pi}_{i}^{p}\right] \qquad \rho_{i}=\frac{\hat{E}_{i}^{p}\rho\hat{E}_{i}^{p\dagger}}{p_{i}},  \label{9}%
\end{equation}
with $i=\left\{  s,f\right\}$.

Let us remark that the Kraus operator corresponding to the successful amplification in Eq.~\eqref{6} is the optimal quantum operation found in Ref.~\cite{Combes}.
On the other hand, the failure Kraus operator in Eq.~\eqref{7} is not uniquely determined a priori, since the only constraint is $ \hat{E}_{s}^{p}{}^\dag \hat{E}_{s}^{p} + \hat{E}_{f}^{p}{}^\dag \hat{E}_{f}^{p} = \mathbbm{1}$.
The Kraus operators reported in Eqs.~\eqref{6} and~\eqref{7} correspond to the so called L\"uders instrument~\cite{Heinosaari2011a} for the POVM $\left\{ \hat{\Pi}_{s}^{p}, \hat{\Pi}_{f}^{p} \right\}$.
Different instruments compatible with the same POVM are obtained by applying outcome-dependent control operations (unitaries or, more generally, CPT maps) 
to a L\"uders instrument. In other words, we focus attention 
to \emph{bare} measurements, where no control operations are applied. In turn, 
this approach is justified by the fact that additional transformations would
not change the gain of the NLA, i.e. they do not add any additional information.

The action of NLA on a generic pure state $|\psi \rangle = \sum_n c_n |n \rangle$.
may be expressed as follows. The successfully amplified state is given by
\begin{equation}
\left\vert \psi_{s}\right\rangle =\frac{\hat{E}_{s}^{p}\left\vert
	\psi\right\rangle }{\sqrt{p_{s}}}=\frac{1}{\sqrt{p_{s}}}\left(  \sum
\limits_{n=0}^{p}g^{n-p}c_{n}\left\vert n\right\rangle +\sum\limits_{n=p+1}%
^{\infty}c_{n}\left\vert n\right\rangle \right)  , \label{11}%
\end{equation}
with%
\begin{equation}
p_{s}=\left[  \sum\limits_{n=0}^{p}g^{2(n-p)}\left\vert c_{n}\right\vert
^{2}+\sum\limits_{n=p+1}^{\infty}\left\vert c_{n}\right\vert ^{2}\right]\,,
\label{12}
\end{equation}
whereas a failed amplification corresponds to a distorted state, given by%
\begin{equation}
\left\vert \psi_{f}\right\rangle =\frac{\hat{E}_{f}^{p}\left\vert
	\psi\right\rangle }{\sqrt{p_{f}}}=\frac{1}{\sqrt{p_{f}}}\left(  \sum
\limits_{n=0}^{p}\sqrt{1-g^{2(n-p)}}c_{n}\left\vert n\right\rangle \right)  ,
\label{14}%
\end{equation}
where
\begin{equation}
p_{f}=\left[  \sum\limits_{n=0}^{p}(1-g^{2(n-p)})\left\vert c_{n}\right\vert
^{2}\right]  . \label{15}%
\end{equation}

The desired action of the NLA is obtained by discarding the distorted output state corresponding to the measurement outcome $f$.
This introduces a trade-off between the degree of amplification and the probability of success.
Indeed, high degrees of amplification, i.e. great values of $g$ and $p$, are achieved at the expense of smaller values of the success probability.
For our purposes it is beneficial to retain also the distorted state, since it still depends on the parameter of interest.

\subsection{Measurement model of the non-deterministic NLA}

It is useful to be more explicit in the realization of the NLA and consider the actual measurement model, instead of the Kraus operators.
In this picture, the action of the NLA is obtained by coupling the input system with an ancillary system (the so-called meter or measuring device), followed by a projective measurement on the latter.
Since there are only two outcomes, we consider a two-level system with its orthonormal basis $\left\{  \left\vert s\right\rangle ,\left\vert f\right\rangle \right\}  $, where $\left\vert s\right\rangle $ is the state of the measuring device when a successful amplification occurs and $\left\vert f\right\rangle $ when the amplification fails.
We assume that the measuring device is prepared in the state $\left\vert f\right\rangle $ before the interaction.
The unitary operator describing the interaction is constructed as follows%
\begin{equation}
\hat{U}_g=\hat{E}_{s}^{p}\otimes\left\vert s\right\rangle \left\langle
f\right\vert +\hat{E}_{f}^{p}\otimes\left\vert f\right\rangle \left\langle
f\right\vert +\hat{A}\otimes\left\vert f\right\rangle \left\langle
s\right\vert +\hat{B}\otimes\left\vert s\right\rangle \left\langle
s\right\vert , \label{16}%
\end{equation}
so that $\hat{E}_{s}^{p} = \langle s \vert \hat U_g \vert f \rangle$ and $\hat{E}_{f}^{p}=\langle f \vert \hat U_g \vert f \rangle$; to ensure the unitarity of $\hat{U}$ we can set $\hat{A} = -\hat{E}_{s}^{p} $ and $\hat{B} = \hat{E}_{f}^{p}$, as shown in~\cite{McMahon}.

The action of the unitary transformation $\hat{U}$ on a generic pure state $|\psi\rangle$ coupled to a measuring device pre-selected in the state $\left\vert
f\right\rangle$ gives rise to the following global state of system and ancilla%
\begin{equation}
\label{eq:FullPureNLA}
\vert \Psi_\mathsf{NLA} \rangle = \hat U_g \left( | \psi \rangle \otimes | f \rangle \right) = \sqrt{p_s} | \psi_s \rangle \otimes |s \rangle + \sqrt{p_f} |\psi_f \rangle \otimes |f\rangle 
\end{equation}
where $\left\vert \psi_{s}\right\rangle $ and $\left\vert \psi_{f}%
\right\rangle $ are respectively the amplified and degraded states appearing in Eqs.~\eqref{11} and~\eqref{14}.

The global state~\eqref{eq:FullPureNLA} is pure since no information has been discarded; this will be useful to set an ultimate bound the the precision of the estimation.
However, in practice such pure state is usually not available and in particular it is useful to consider its \emph{decohered} version~\cite{Combes2014}
\begin{equation}
\rho_\mathsf{NLA} = p_{s}\left\vert \psi_{s}\right\rangle \left\langle \psi
_{s}\right\vert \otimes\left\vert s \right\rangle \left\langle s\right\vert
+p_{f}\left\vert \psi_{f}\right\rangle \left\langle \psi_{f}\right\vert
\otimes\left\vert f\right\rangle \left\langle f\right\vert . \label{eq:rhoNLA}%
\end{equation}
This mixed state can be thought as a state where the ancilla is used only to store the classical outcomes of the measurement.
In general, this state contains less information than the pure state $|\Psi_\mathsf{NLA} \rangle$, but we will see that its information content pertains to a more realistic metrological scheme.
Furthermore, we are going to show that for this particular estimation problem the two states contain the same amount of information about the parameter.

\section{Cramér-Rao bounds for the estimation of the gain}
\label{s:CRB}
We study three different strategies to infer the value of the gain of a non-deterministic NLA with squeezed vacuum and coherent states used as probes; we assess their performances in term of their respective QCRBs.

\subsection{The three schemes}
\subsubsection{Global state}
The first strategy we consider is a sequential measurement scheme, where we first measure the system indirectly via the POVM $\{ \hat{\Pi}_{s}^{p}, \hat{\Pi}_{f}^{p} \}$ and then we perform a final measurement on the conditional states of the system.
By assuming to be able to perform any measurement on the conditional state the correct figure of merit is the \emph{effective} QFI (we adopt the terminology introduced in~\cite{Albarelli}):
\begin{equation}
\mathcal{Q}_\mathsf{eff}(g)=p_{s} \mathcal{Q}_{s}(g)+p_{f} \mathcal{Q}_{f}(g)+\mathcal{F}_{c}\left(  \left\{
p_{s},p_{f}\right\}  \right)  , \label{eq:effQFI}%
\end{equation}
where $\mathcal{Q}_{s}(g)=\mathcal{Q}\left( | \psi_s \rangle \right)$ and $\mathcal{Q}_{f}(g)=\mathcal{Q}\left( | \psi_f \rangle \right)$ are respectively the QFI of the amplified
state and the degraded one, while $\mathcal{F}_{c}$ is the classical FI associated with the distribution probability $\left\{  p_{s},p_{f}\right\}$.
We are going to study this sequential strategy more in detail in Sec.~\ref{s:FI}.

Let us show that this quantity corresponds to the QFI of the state $\rho_\mathsf{NLA}$ defined in~(\ref{eq:rhoNLA}), see also~\cite{Combes2014,Shitara2016}.
We resort to the primary definition of the QFI and we evaluate the symmetric logarithmic derivative.
The state $\rho_\mathsf{NLA}$ can be expressed as a $2 \times 2$ block matrix, with diagonal elements $p_{f}\rho_{s}$ and $p_{f}\rho_{f}$ and null extra-diagonal ones.
This particular shape enables for a straightforward evaluation of the SLD leading to%
\begin{equation}
\partial_{g}\rho_\mathsf{NLA}=\frac{1}{2}\left\{  L_{s},p_{s}\rho_{s}\right\}
\otimes\left\vert s\right\rangle \left\langle s\right\vert +\frac{1}%
{2}\left\{  L_{f},p_{f}\rho_{f}\right\}  \otimes\left\vert f\right\rangle
\left\langle f\right\vert , \label{18}%
\end{equation}
where $L_{s,f}=\partial_{g}\ln p_{s,f}+2\partial_{g}\rho_{s,f}$ and $\left\{
,\right\}  $ denotes the anti-commutator. After gathering together the two
terms of the right side, we obtain the equation for the SLD of the overall state $\rho_\mathsf{NLA}$ as 
\begin{equation}
\partial_{g}\rho_\mathsf{NLA}=\frac{1}{2}\left\{  L,\rho_\mathsf{NLA}\right\}  ; \label{19}%
\end{equation}
where $L$ is a $2 \times 2$ block matrix with $L_{s}$ and $L_{f}$ on the diagonal and null off-diagonal elements.
The final expression of the QFI is then carried out using its well-known definition $\mathrm{Tr}\left[  \rho_\mathsf{NLA}L^{2}\right]  $ and gives Eq.~\eqref{eq:effQFI}.

We notice that the sum of the two first terms in Eq.~\eqref{eq:effQFI} represents the average QFI of the two pure states w.r.t. the probability distribution $\left\{  p_{s},p_{f}\right\}$.
The expressions are found to be%
\begin{eqnarray}
\mathcal{Q}_{s}(g)=-\left(  \frac{\partial_{g}p_{s}}{p_{s}}\right)  ^{2}+\frac{4}{p_{s}%
}\sum\limits_{n=0}^{p-1}(n-p)^{2}g^{2(n-p-1)}\left\vert c_{n}\right\vert ^{2},
\label{21} \\
\mathcal{Q}_{f}(g)=-\left(  \frac{\partial_{g}p_{f}}{p_{f}}\right)  ^{2}+\frac{4}{p_{f}%
}\sum\limits_{n=0}^{p-1}\frac{(n-p)^{2}g^{4(n-p)-2}}{(1-g^{2(n-p)})}\left\vert
c_{n}\right\vert ^{2}, \label{22}%
\end{eqnarray}
where $\partial_{g}p_{s,f}$ are the derivatives of the probabilities $p_{s,f}$
with respect to $g$ (see~\ref{app:effQFI} for more details).
The FI of the classical probability distribution is
\begin{equation}
\mathcal{F}_{c}\left(  \left\{  p_{s},p_{f}\right\}  \right)  =
\frac{\left( \partial_{g}p_{s} \right)  ^{2}}{p_{s}}+  \frac{\left( \partial_{g}p_{f}%
\right)^{2} }{p_{f}}. \label{23}%
\end{equation}
When summing up all the terms in Eq.~\eqref{eq:effQFI}, we see that $\mathcal{F}_{c}\left(  \left\{  p_{s},p_{f}\right\}  \right)  $ cancels out the the first ``classical'' terms in Eqs.(\ref{21},\ref{22}), more details are provided in~\ref{app:effQFI}.

Due to the fact that the NLA only changes the amplitudes of the Fock components of a quantum state but does not add relative phases between such components, we have that the normalized states are orthogonal to their derivatives, i.e. $\langle \psi_s | \partial_g \psi_s \rangle=\langle \psi_f | \partial_g \psi_f \rangle=0$~\footnote{This property is not true in general. For a generic dependence on the parameter $\lambda$ a complex state vector only satisfies $\mathrm{Re}\langle \psi | \partial_\lambda \psi \rangle = 0$.
The full orthogonality condition makes the quantum case similar to the classical case of real valued probability distributions (i.e. real valued normalized vectors).}.
From this identity it easy to prove that the QFI of the state $|\Psi_\mathsf{NLA} \rangle$ defined in~\eref{eq:FullPureNLA} is equal to the effective QFI, see details in~\ref{app:genericMeter}.
To sum up, we have found the following equalities:
\begin{equation}
\mathcal{Q}_\mathsf{eff}(g) = \mathcal{Q} \left( \rho_{\mathsf{NLA}} \right) = \mathcal{Q} \left( | \Psi_\mathsf{NLA} \rangle \right) , \label{eq:Qeqs}
\end{equation}
where the two states are defined in Eqs.~\eqref{eq:FullPureNLA} and~\eqref{eq:rhoNLA}.
Eq.~\eqref{eq:Qeqs} shows that having full access to the global system plus ancilla state and being able to perform arbitrary measurements (e.g. projections onto entangled states) is not useful.
The sequential scheme we described, measuring the ancilla first and the system afterwards, is indeed optimal.

We notice that by implementing the global unitary~\eqref{16} in full generality one could be able to obtain more information about the parameter $g$, since such an operation could be applied to arbitrary (entangled) states of the bosonic mode plus the ancillary qubit and not only on the state $| \psi \rangle \otimes |f \rangle$.
We remark that our approach is to treat the NLA as given device to calibrate, thus we also consider the preparation of the initial state $|f \rangle$ as built into the operation of the device.
However, for completeness in~\ref{app:genericMeter} we consider the simplest scheme: a separable input state, but with an arbitrary state of the meter qubit, i.e. $ |\psi \rangle \otimes \left( \alpha | s \rangle + \beta | f \rangle \right)$.
We show that this approach never yields more information about the parameter than preselecting the meter state $|f \rangle$.

\subsubsection{Unconditional state}
In the second scenario, we consider the unconditional state arising from the action of the NLA on the probe states.
Its expression is derived by tracing out the measuring device (M) in Eq.~(\ref{eq:rhoNLA}) and reads%
\begin{equation}
\rho_\mathsf{unc}=\mathrm{Tr}_{M}\left[  \rho_\mathsf{NLA}\right] = 
p_s \left\vert \psi_{s}\right\rangle \left\langle \psi_{s}\right\vert +
p_f \left\vert \psi_{f}\right\rangle
\left\langle \psi_{f}\right\vert . \label{eq:RhoUncond}%
\end{equation}
We notice that the resulting state is a mixture and the evaluation of the QFI is carried out numerically after expanding the amplified and degraded states in the Fock basis.

The figure of merit to assess this scheme is the QFI of the state $\rho_\mathsf{unc}$, denoted as $\mathcal{Q}_\mathsf{unc} (g)$.
In general the information obtainable by considering this mixed state is less than the effective QFI we previously introduced.
This can be easily understood in terms of the monotonicity properties of the QFI~\cite{Petz2010}, since the partial trace is a completely positive and trace preserving map.
Therefore, we have the inequality
\begin{equation}
\mathcal{Q}_\mathsf{unc} (g) \leq \mathcal{Q}_\mathsf{eff}(g) \;
\end{equation}
which is also known as the \emph{extended convexity} of the QFI~\cite{Alipour,Ng2016}.

\subsubsection{Successfully amplified state}
In the previous schemes, we took into consideration the contributions of both the states in the mixture.
A widespread approach is to focus only on the relevant states which concentrate information on the unknown parameter~\cite{Gendra,Gendra2,Calsamiglia,Walk}.
In this spirit we also study the effect of only taking into account the contribution of the amplified states, discarding the distorted ones.
This represent the third and last strategy we consider.

Before proceeding, we stress the fact that the QFI of the post-selected amplified generic state may be larger than that of the overall state.
According to this observation, one may expect to attain a better sensitivity by considering only the amplified states.
Nevertheless, as seen before, the QFI by itself is not the relevant quantity for the estimation accuracy but the actual bound on the variance is the CRB.
The main trouble with this post-selection scheme is that the estimator is built with a smaller sample since the distorted states are discarded, while the other proposed schemes make use of
all the available probes.
Thereby, to fairly compare the different schemes we should consider these quantities: $M_{s} \mathcal{Q}_{s}(g),$ $M \mathcal{Q}_\mathsf{eff}(g),$ and $ M \mathcal{Q}_\mathsf{un}(g)$, where $M_{s}$ is the number of measurements performed on the post-selected amplified states while $M$ refers to the whole sample~\cite{Tanaka}.

We notice that considering the general case without any assumptions, this issue cannot be readily fixed.
Here, we consider the asymptotic case of infinite runs where the CRB can
effectively be saturated.
The number of measurement involved in this third strategy is thus given by $M_{s} = M p_{s}$ and the correct figue of merit for this estimation scheme is the QFI rescaled by the success probability $p_{s}\mathcal{Q}_{s}(g)$, see also a similar discussion in~\cite{Combes2014}.
For this quantity we have the inequality
\begin{equation}
p_s \mathcal{Q}_s (g) \leq \mathcal{Q}_\mathsf{eff}(g) \;,
\end{equation}
which follows from the definition of the effective QFI.
\subsection{Results and discussions}
\begin{figure}[ptb]
		\includegraphics[width=.45\textwidth]{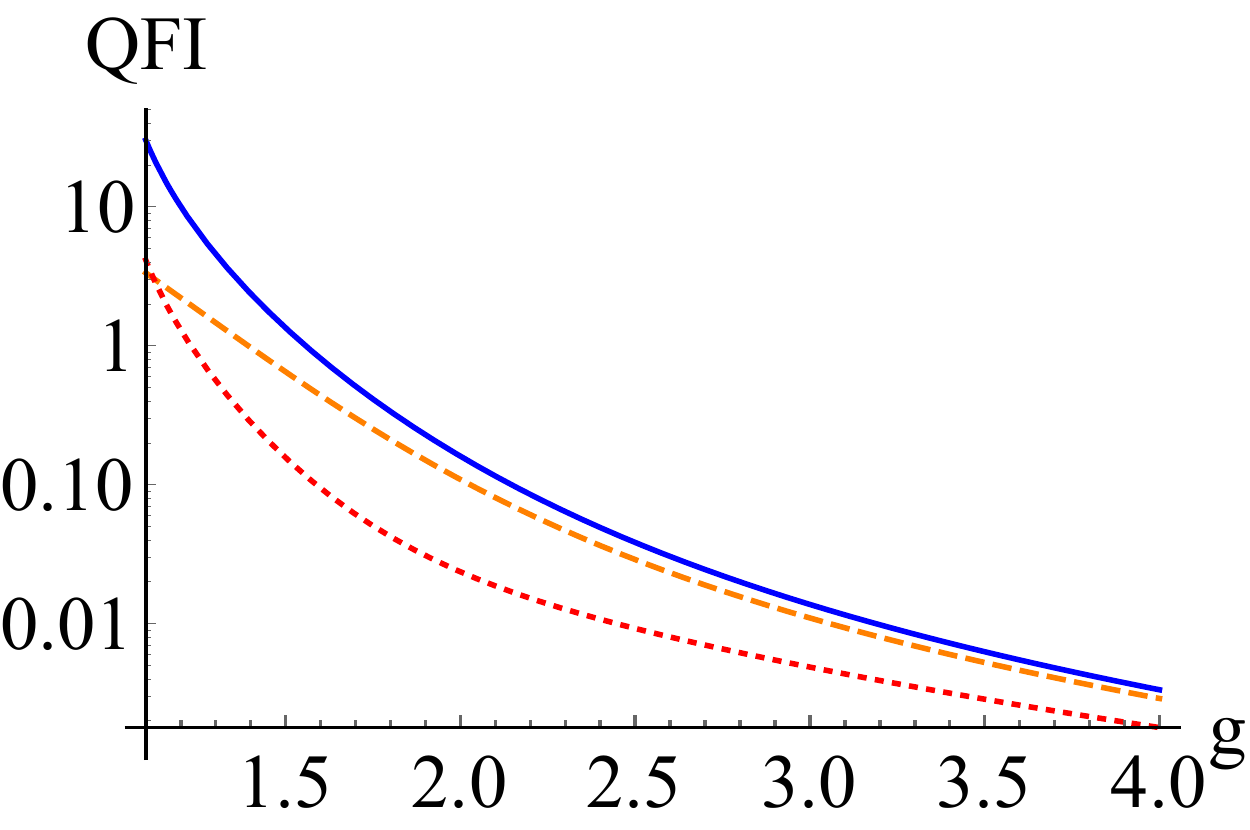}
	\hfill
		\includegraphics[width=.45\textwidth]{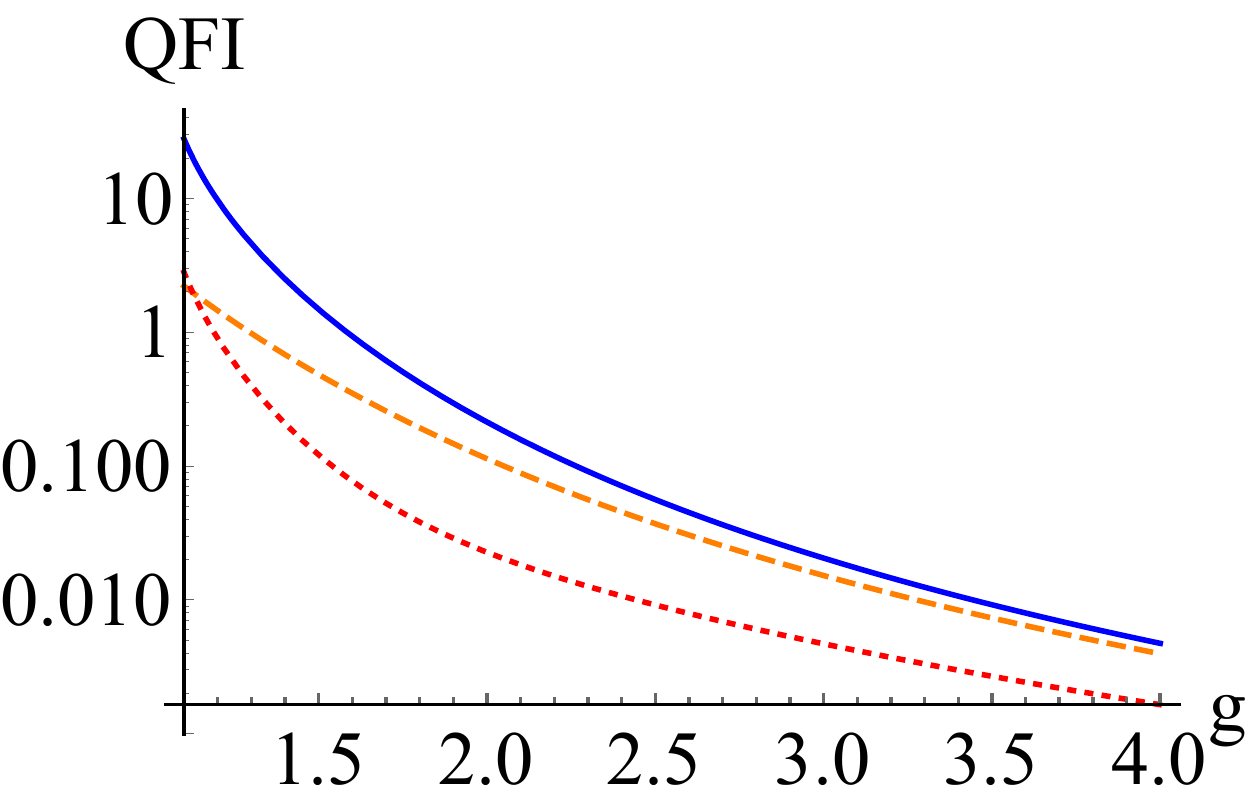}
	\caption{Plots of the QFI of the proposed strategies as functions of $g$ at fixed input energy $\bar{n}=1$ and threshold $p=3.$ The blue solid line represents the effective QFI $\mathcal{Q}_\mathsf{eff}$, the orange dashed line denotes $p_{s}\mathcal{Q}_{s}(g)$ and the red dotted line represents the QFI of the unconditional state $\mathcal{Q}_\mathsf{unc}$.
	The left panel shows results for a squeezed vacuum state; the right one for a coherent input state.}%
	\label{fig:diffStrategies}
\end{figure}
We now proceed to compare the performance of these strategies in terms of the figures of merit we have just introduced.
In particular we focus on two categories of probes: coherent input states and squeezed vacuum.
Coherent states are defined by the coefficients $c_{n}=e^{-|\alpha|^2} \frac{\alpha^{n}}{\sqrt{n!}}$ and have an average number of photons $\bar{n}=|\alpha|^2$, while the squeezed vacuum corresponds to $c_{2n}= \frac{1}{\sqrt{\mu}}\left(  \frac{\nu}{2\mu}\right)^n \frac{\sqrt{(2 n)!}}{n!}$ and $c_{2n+1}=0$ with $\nu = \sinh r$ and $\mu = \cosh r $ (we choose a real squeezing parameter $r$), the mean photon number is $\bar{n} = \nu^2$.

\begin{figure}[ptb]
		\includegraphics[width=.45\textwidth]{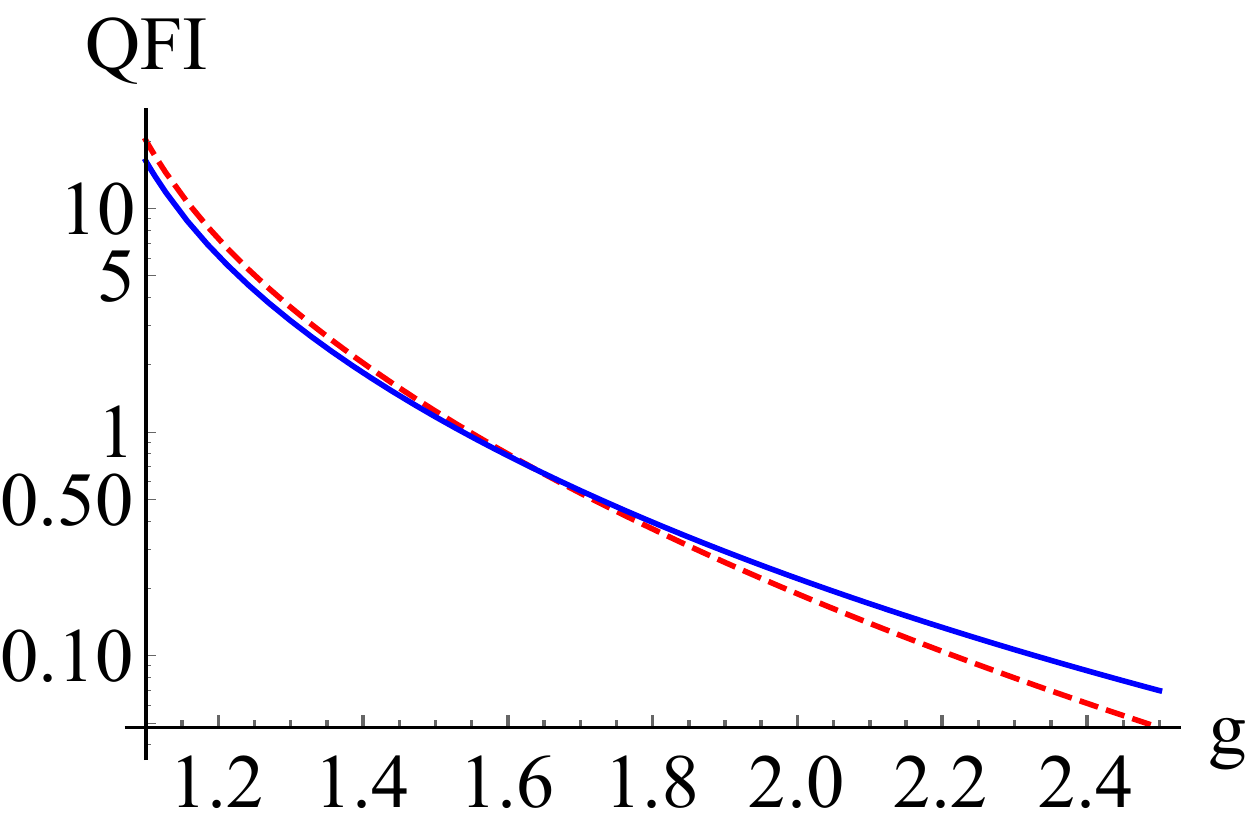}
	\hfill
		\includegraphics[width=.45\textwidth]{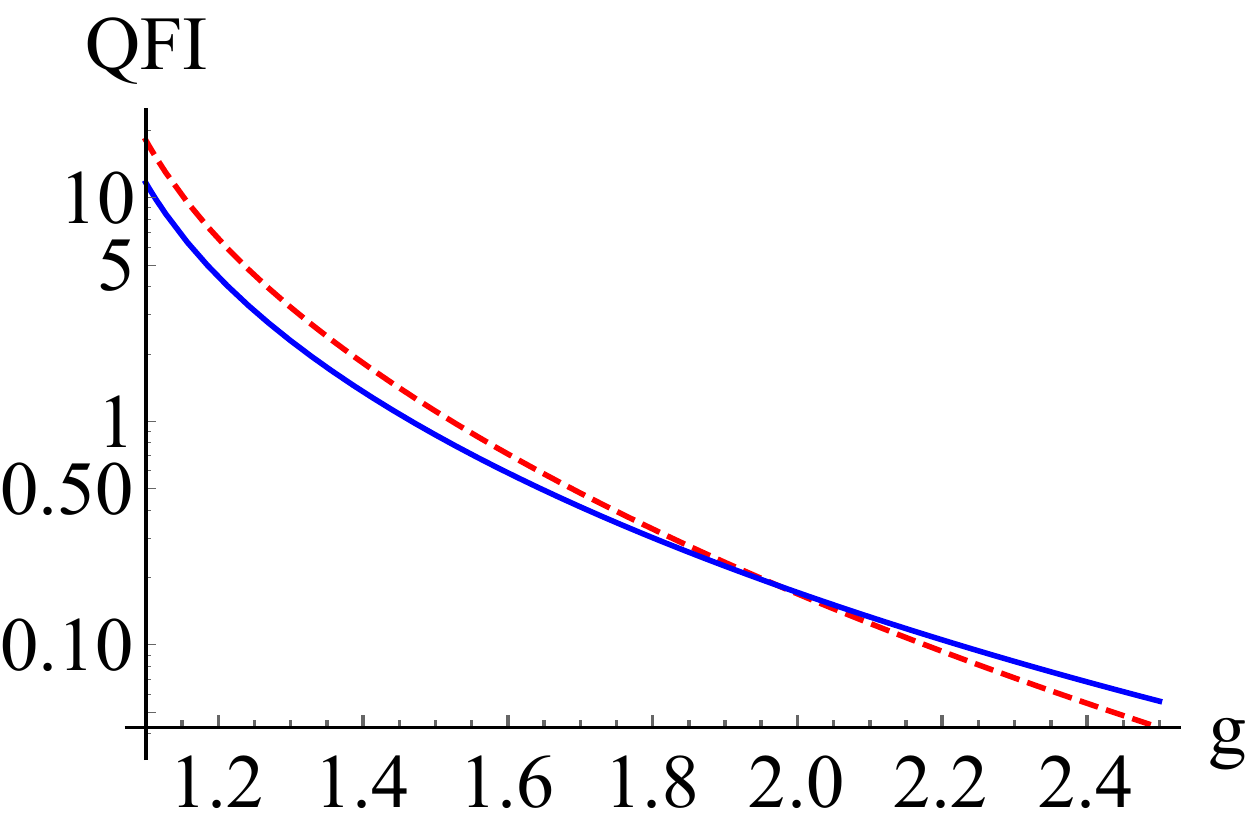}
	\caption{ Plots of the effective QFI as a function of the gain for different
		probe states at fixed value of the truncation order (p=2) and distinct mean
		input energy values. The red dotted curves denote a squeezed vacuum state, while the
		blue solid ones represent a coherent input state.
		The left panel shows results for $\overline{n}=1$; the right one for $\overline
		{n}=1.5$. }%
		\label{fig:effQFIdiffstates}
\end{figure}

\begin{figure}[ptb]
	\centering
	\begin{subfigure}[b]{0.475\textwidth}
		\centering
		\includegraphics[width=\textwidth]{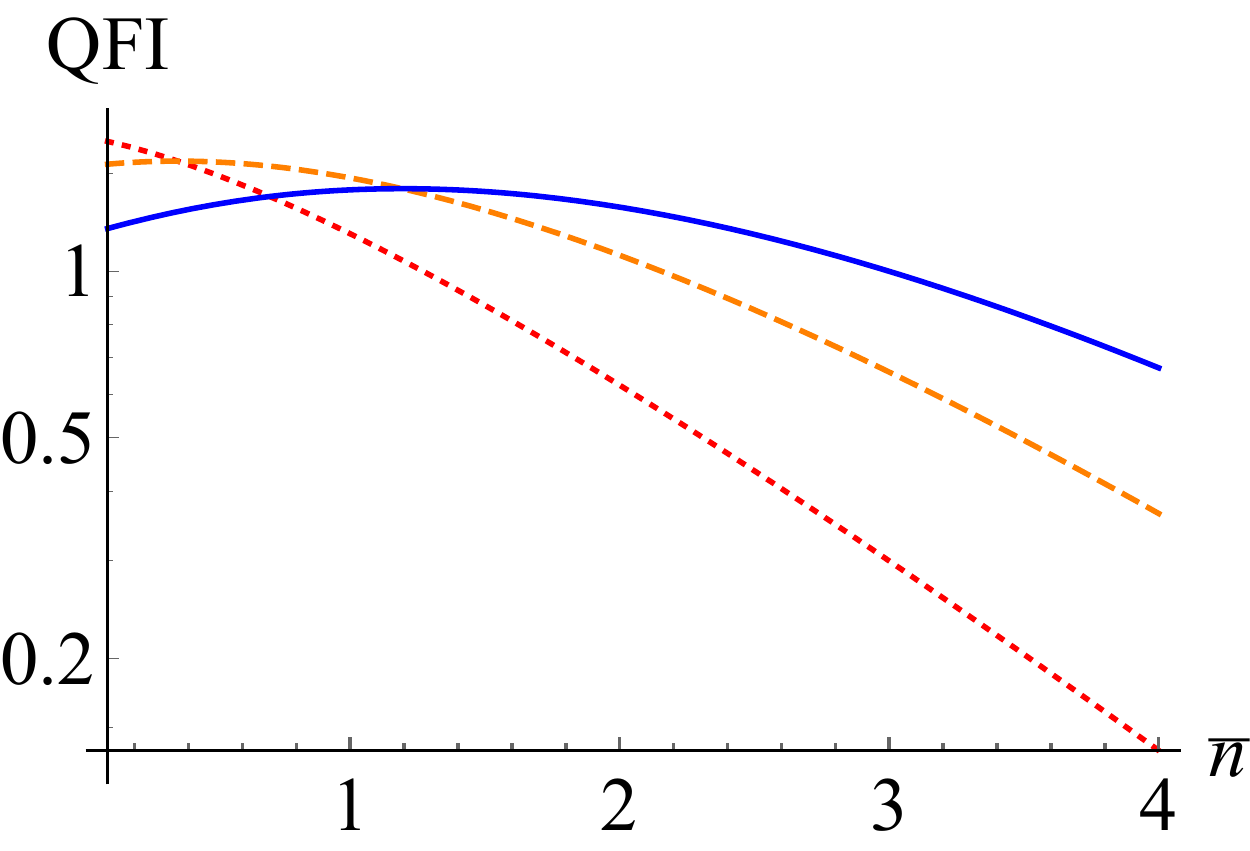}
		\caption[Network2]%
		{{\small}}
		\label{}
	\end{subfigure}
	\hfill\begin{subfigure}[b]{0.475\textwidth}
		\centering
		\includegraphics[width=\textwidth]{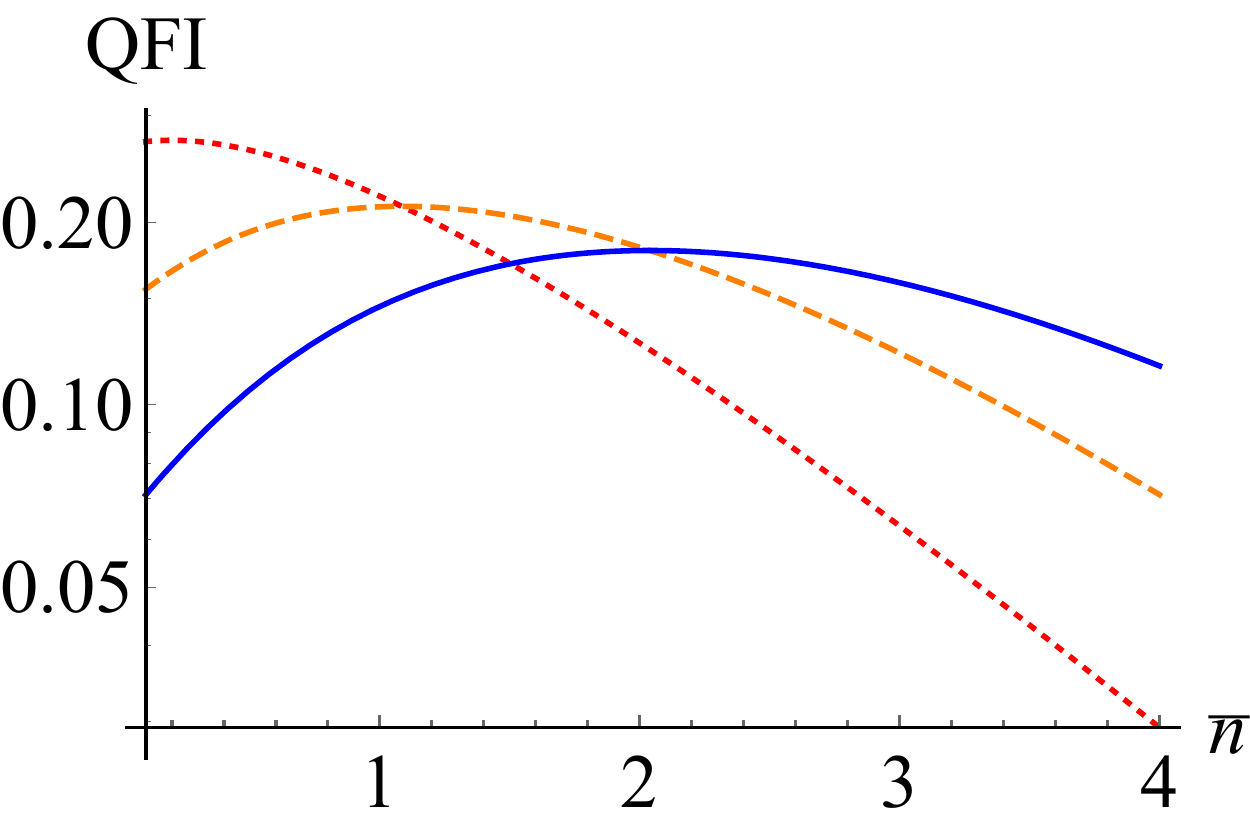}
		\caption[]%
		{{\small}}
		\label{}
	\end{subfigure}
	\vskip\baselineskip
	\begin{subfigure}[b]{0.475\textwidth}
		\centering
		\includegraphics[width=\textwidth]{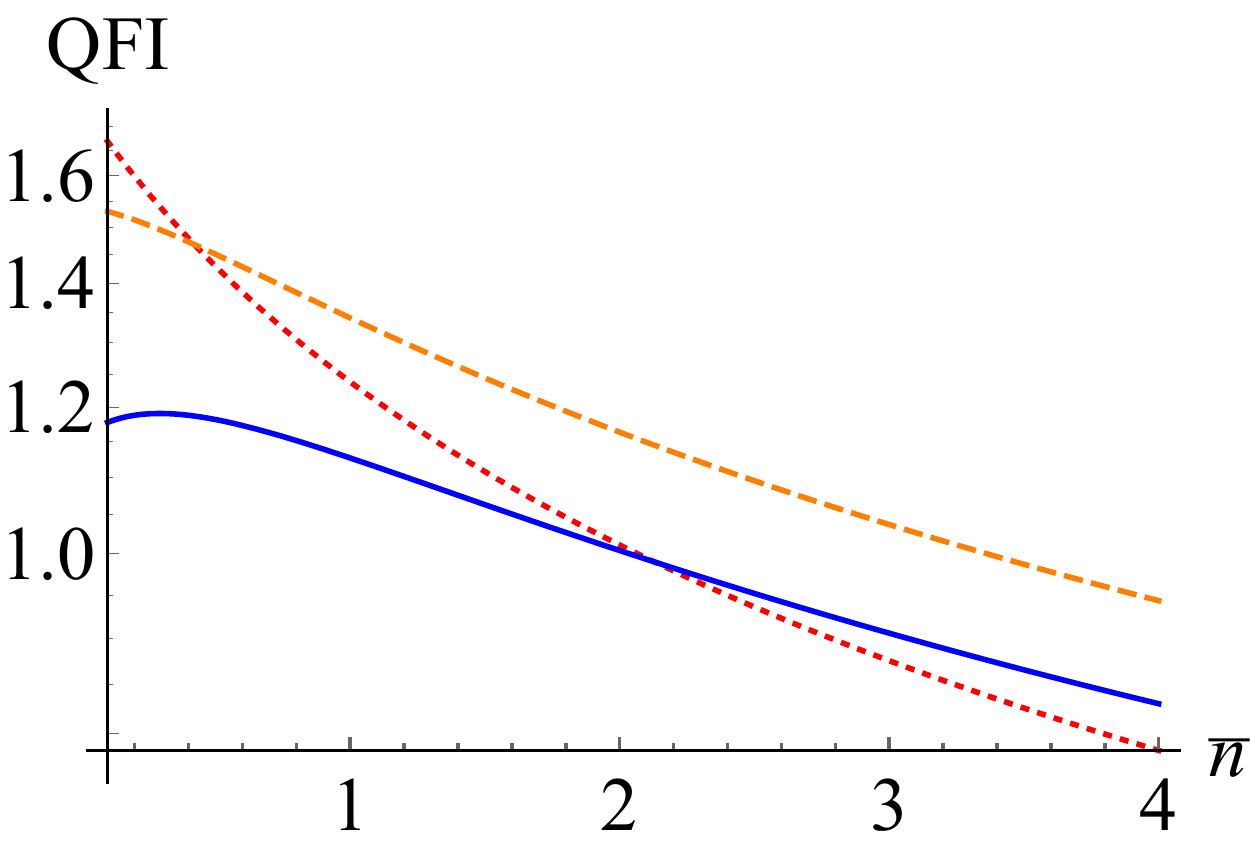}
		\caption[]%
		{{\small}}
		\label{}
	\end{subfigure}
	\quad\begin{subfigure}[b]{0.475\textwidth}
		\centering
		\includegraphics[width=\textwidth]{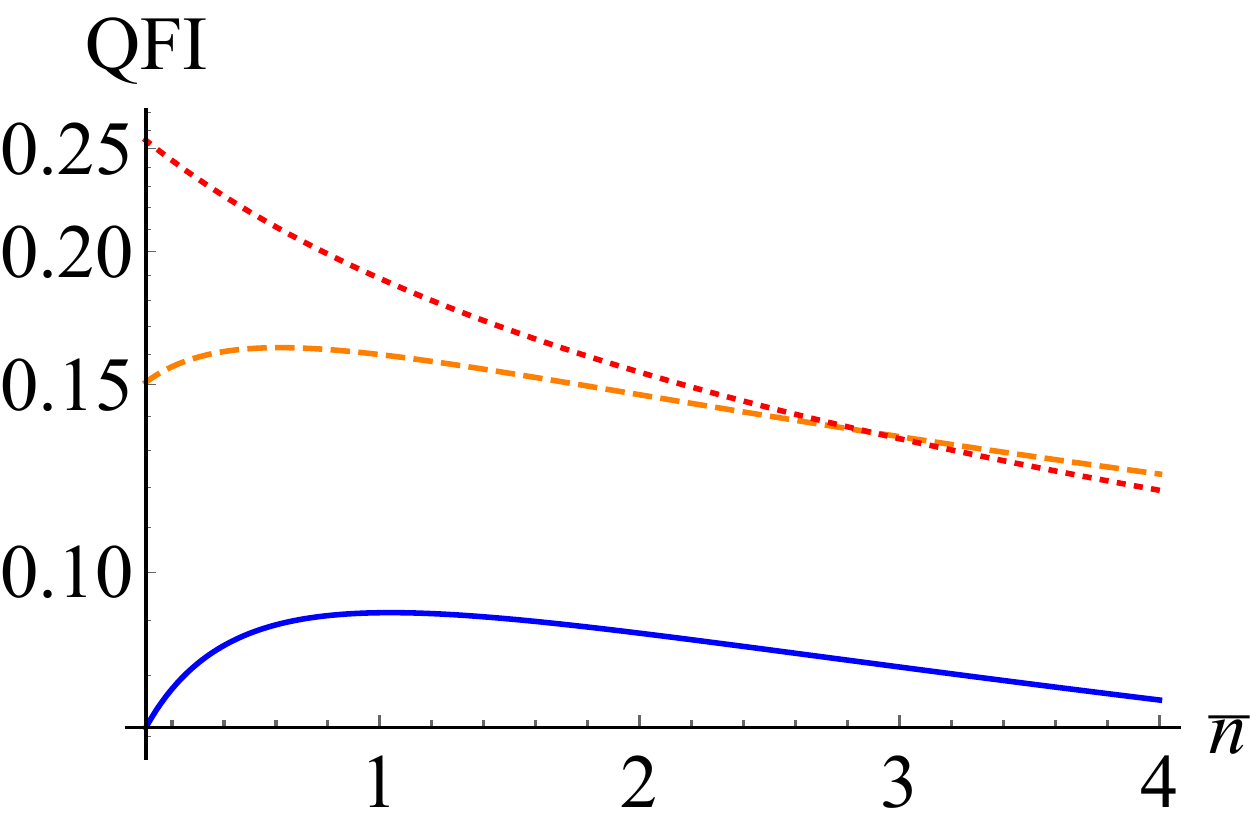}
		\caption[]%
		{{\small }}
		\label{}
	\end{subfigure}
	\caption{{\protect\small Plots of the effective QFIs as
			functions of the mean photons number $\bar{n}$. The panels (a) and (b) on top show $\mathcal{Q}_\mathsf{eff}$ for a coherent state, considering
			different values of the threshold (the red dotted line denotes p=2, the orange dashed line
			p=3 and the blue solid line p=4).
			In the left panel (a) we have $g=1.5$, while in panel (b) we have $g=2$.
			The two bottom panels (c) and (d) show the same quantities with squeezed vacuum probe, for the same values of the other parameters. }}%
	\label{fig:effQFIvsNphot}%
\end{figure}

In Fig.~\ref{fig:diffStrategies} we show that all the figures of merit are decreasing functions of the gain and are very close to zero for gain values exceeding $4$, regardless of the considered probe.
In addition, we clearly notice that considering the sequential scheme characterized by $\mathcal{Q}_\mathsf{eff}(g)$ offers the most accurate estimate for the gain for both the coherent input state and squeezed vacuum, in line with the inequalities we presented.
As previously noticed, this shows that the apparent enhancement in the information obtained from
post-selecting only the amplified states is cancelled out by the small probability of success~\cite{Tanaka}.
However, we also find that, except for values of $g$ close to $1$, the precision of the design based on post-selecting the successfully amplified state still gives a better precision than the unconditional state, regardless of the considered probe state.
To sum up our results, under the assumptions of weak mean input energies and values of the gain $g$ exceeding 1.2, we found a hierarchy between the different strategies under study
\begin{equation}
\mathcal{Q}_\mathsf{eff}(g)>p_{s}\mathcal{Q}_{s}(g) \geqslant \mathcal{Q}_\mathsf{un}(g);
\end{equation}
the inequalities between $\mathcal{Q}_\mathsf{eff}$ and the other quantities are expected from the general arguments of the previous section, while the inequality between $p_{s}\mathcal{Q}_{s}(g)$ and $\mathcal{Q}_\mathsf{un}(g)$ holds only under the assumptions mentioned above and is not true in general.

In Fig.~\ref{fig:effQFIdiffstates} we depict the plots of the effective QFI for the two considered classes of probe states, at a fixed truncation order and for different values of the mean input energy.
Our results show that, for relatively weak values of the gain, the squeezed vacuum offers a better sensitivity than a coherent probe while for g exceeding a certain value which depends on the mean input energy, a coherent input state is more efficient. 

Moreover, by using squeezed states we also found an important accuracy enhancement for small values of the gain and greater mean input energies.
These conclusions are shown in Fig~\ref{fig:effQFIvsNphot}.
Indeed, the sub-figures (a) and (c) obtained for $g=1.5$ show an enhancement of the accuracy with the squeezed vacuum for all the considered values of the truncation order $p$ and the input energy.
On the other hand the remaining sub-figures on the right panel (for $g=2$) show that in this regime there is no trivial relationship between the considered parameters.
Summing up, our results allow to choose the Gaussian probe state with the optimal input energy in order to infer the unknown value of the gain assuming a given truncation order.

\section{Extraction of the maximum amount of information via feasible
measurements}
\label{s:FI}
As we have seen in the previous Sections, the QCRB achievable
by measuring the global pure state $| \Psi_\mathsf{NLA} \rangle$ corresponds 
to the best sensitivity to infer the value of the gain. 
We also noticed that this ultimate bound is saturated by the 
sequential strategy, for which the precision is quantified by 
$\mathcal{Q}_\mathsf{eff}(g)$.

In order to investigate the performances of feasible measurements, 
let us briefly review this sequential strategy.
The main idea is to extract the maximum possible information by taking into consideration both the contributions of the amplified and degraded states as well the information coming from the statistics of the post-selection process itself.
The post-selection is achieved by performing a projective measurement on the orthogonal basis vectors of the measuring apparatus leading to a conditional state of the probe.
When a successful amplification is heralded, the input state is transformed in the required way $\left\vert \psi\right\rangle \longmapsto\left\vert \psi_{s}\right\rangle,$ while when the measuring device displays a failure output, the probe state is degraded $\left\vert \psi\right\rangle \longmapsto\left\vert \psi_{f}\right\rangle .$
In both cases, the resulting amplified and distorted states undergo a strong measurement; in particular, we will study photon counting and homodyne detection.
\begin{figure}
	\centering
	\includegraphics[width=0.7\linewidth]{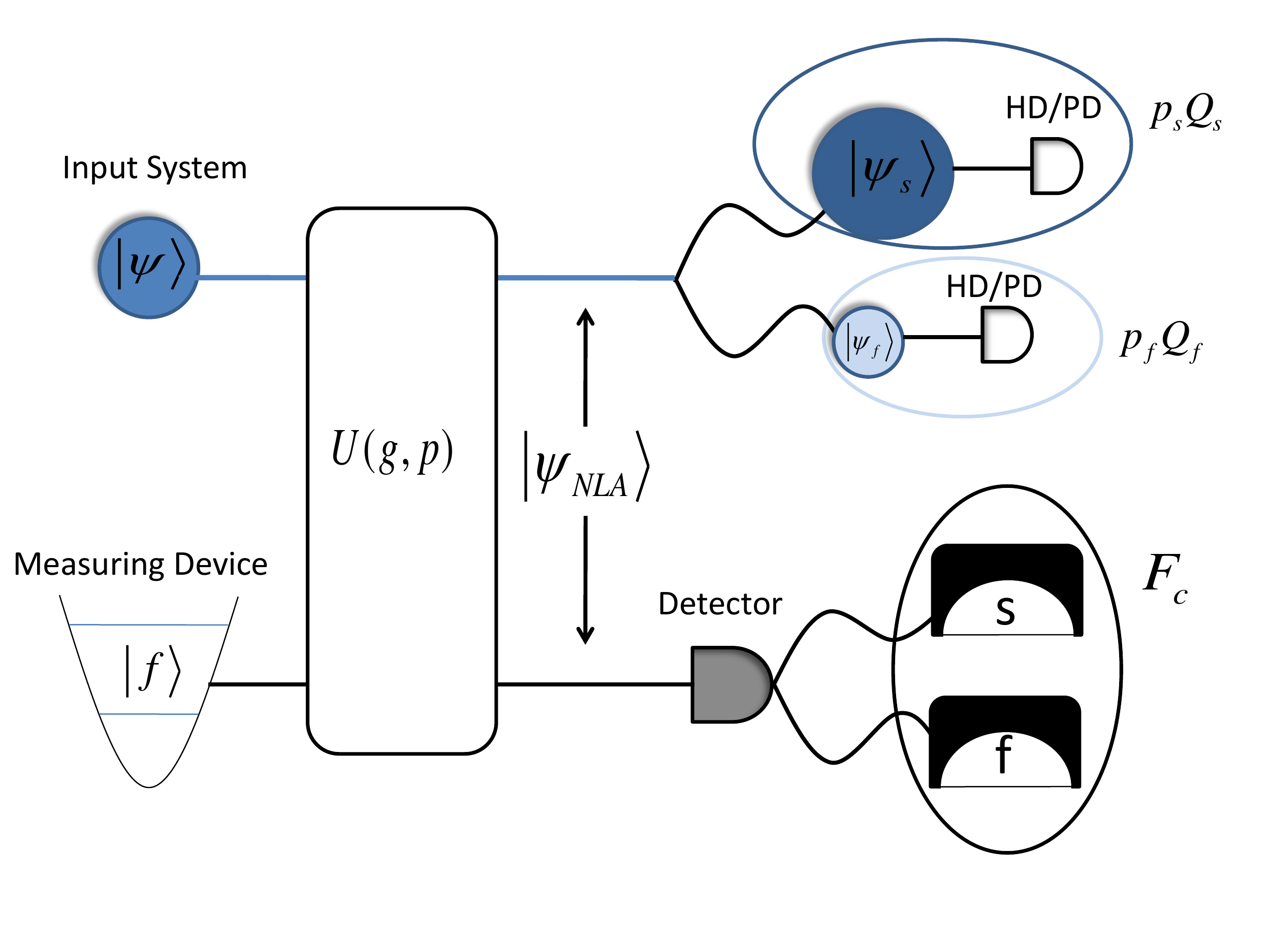}
	\caption{Setup for the proposed measurement scheme. The source of the different contributions for the effective QFI are underlined: the FI of the classical distribution ${p_{s},p_{f}} $ comes from post-selecting one of the measurement device's outcomes whereas the average of the QFI of the amplified and degraded probes is extracted via a strong measurement on these latter (Photo-detection (PD) or Homodyne detection (MD)).}
	\label{fig:schematic-representation}
\end{figure}

Here, we show that the effective QFI corresponds to the classical FI of the sequential measurement scheme, when the optimal measurement on the conditional states are performed.
Moreover we find that these optimal measurements are fixed; they do not depend on the value of $g$ and they are optimal for both conditional states, thereby making the scheme appealing for possible implementation with current technology.
\begin{figure}[ptb]
	\centering
	\begin{subfigure}[b]{0.475\textwidth}
		\centering
		\includegraphics[width=\textwidth]{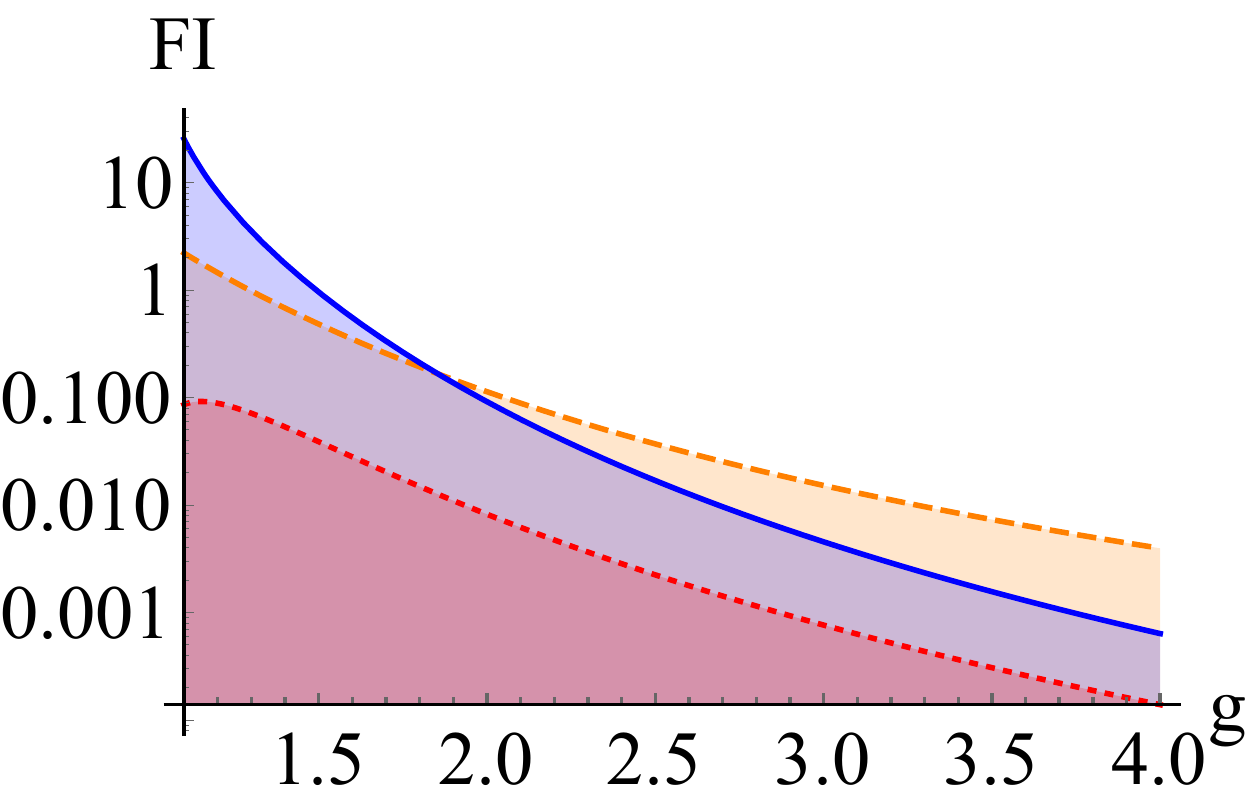}
		\caption[Network2]%
		{{\small}}
		\label{}
	\end{subfigure}
	\hfill\begin{subfigure}[b]{0.475\textwidth}
		\centering
		\includegraphics[width=\textwidth]{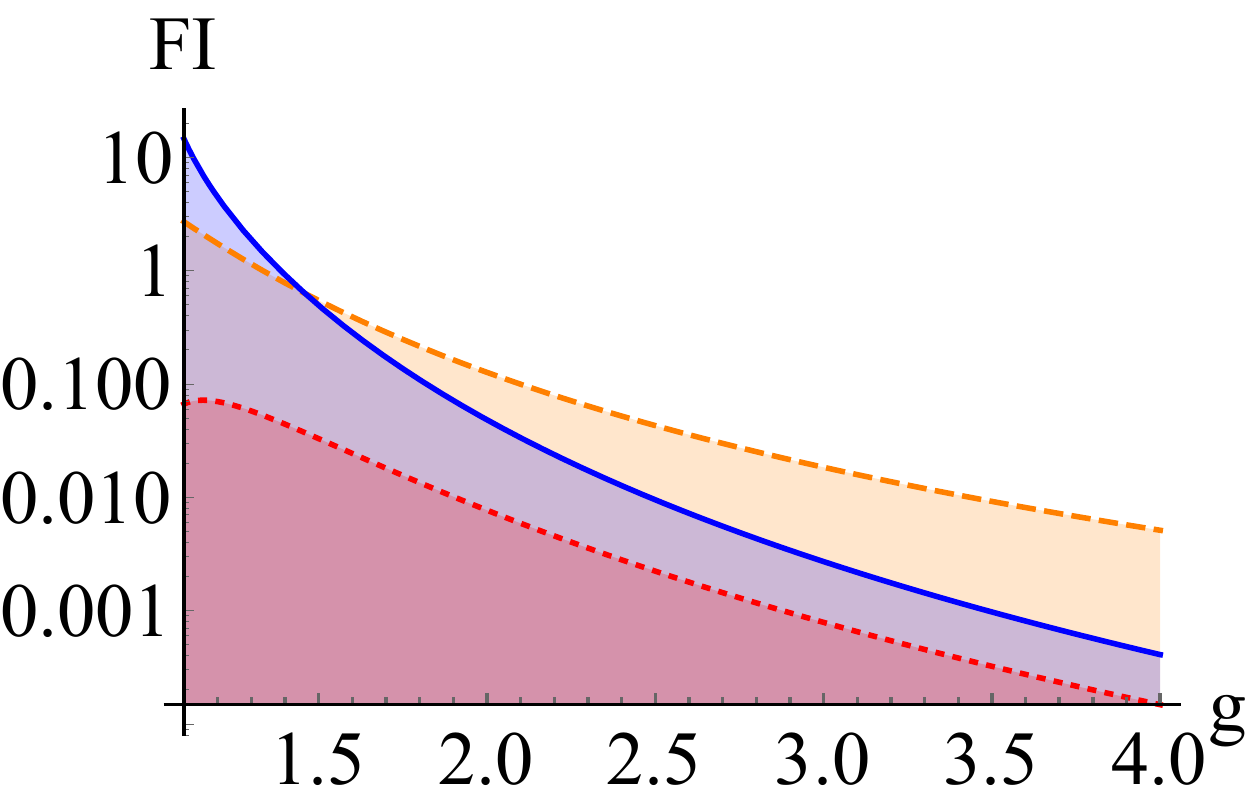}
		\caption[]%
		{{\small}}
		\label{}
	\end{subfigure}
	\vskip\baselineskip
	\begin{subfigure}[b]{0.475\textwidth}
		\centering
		\includegraphics[width=\textwidth]{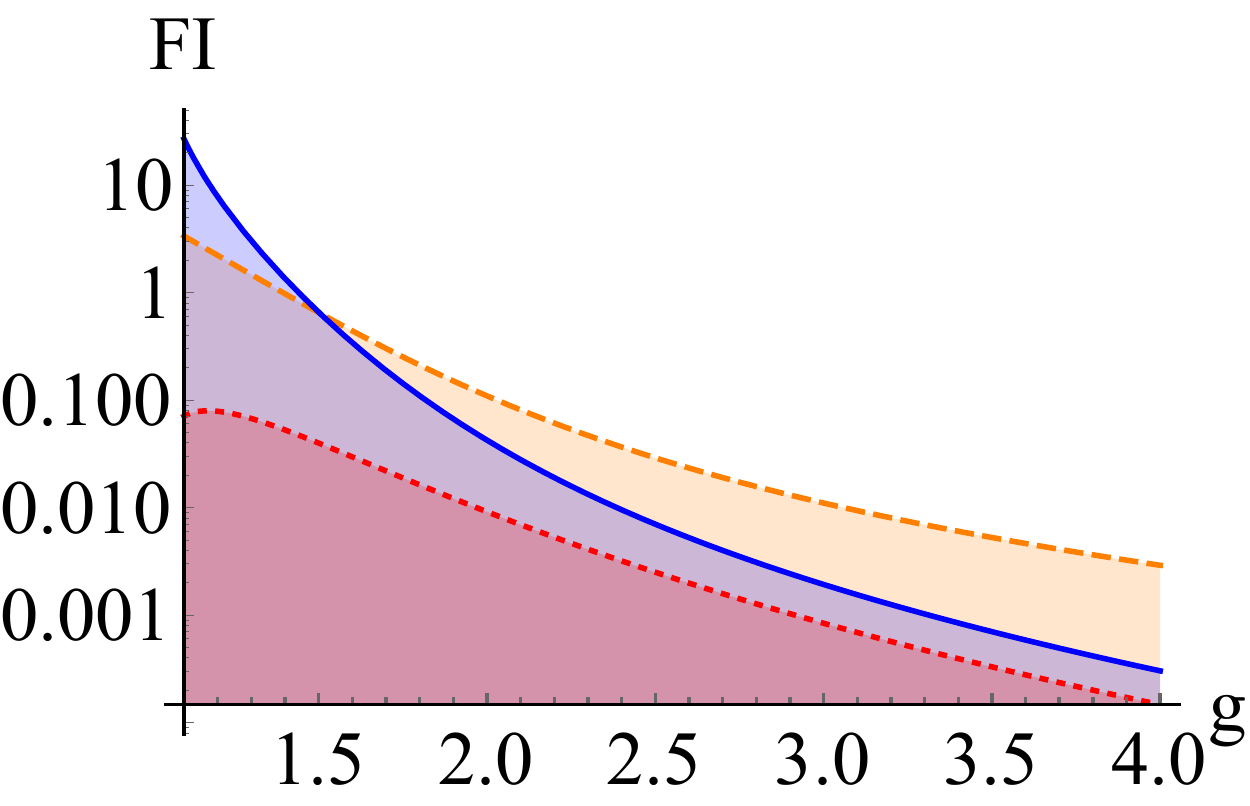}
		\caption[]%
		{{\small}}
		\label{}
	\end{subfigure}
	\quad\begin{subfigure}[b]{0.475\textwidth}
		\centering
		\includegraphics[width=\textwidth]{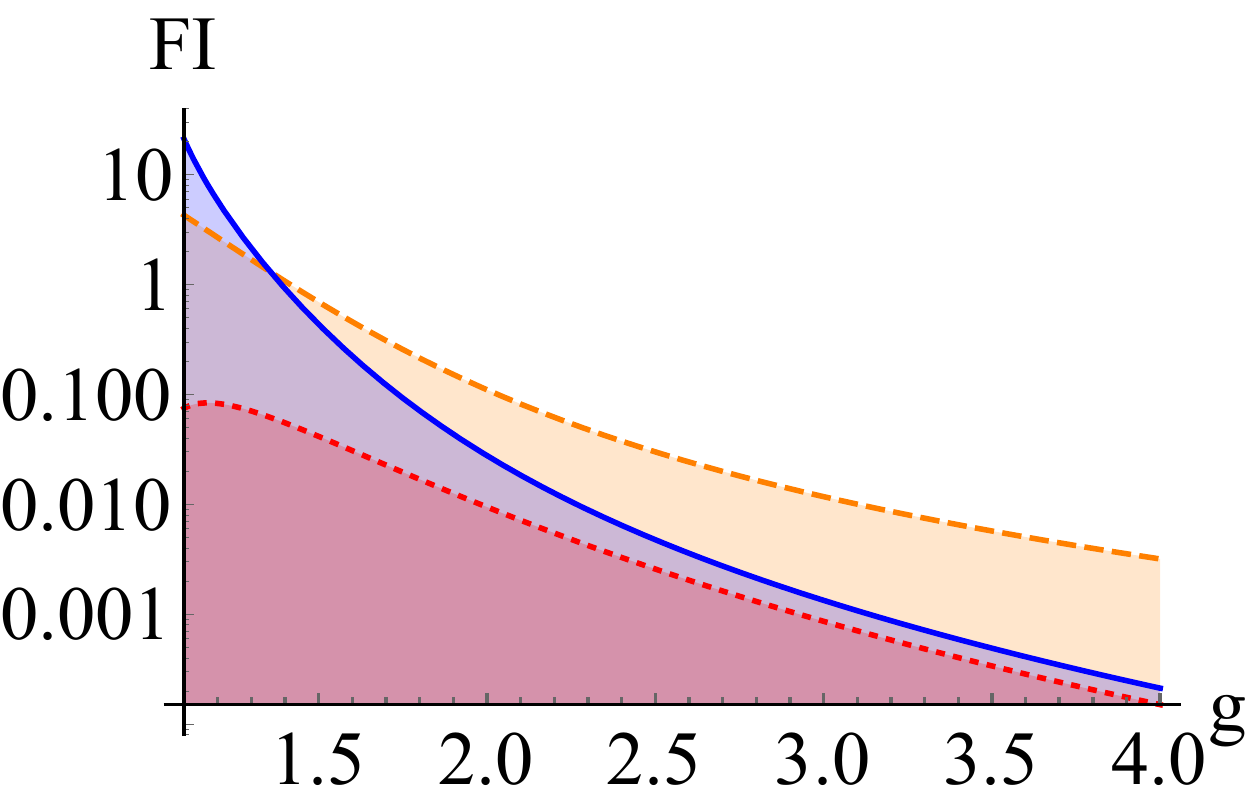}
		\caption[]%
		{{\small }}
		\label{}
	\end{subfigure}
	\caption{{\protect\small Plot of the different contributions to the effective QFI as
			functions of $g$ at fixed threshold $p=3$ for input energies $\overline{n}=1$ in the left panels (a) and (c), $\overline{n}=2$ in right panels (b) and (d). The top panels (a) and (b) are obtained for a coherent input state, the bottom panels (c) and (d) for the squeezed vacuum.
			The blue solid line represents the  contribution of the FI of the distribution $\{p_{s},p_{f}\}$, the orange dashed line denotes the successfully amplified states $p_{s}Q_{s}(g)$ and the red dotted line is representative of the distorted states $p_{f}Q_{f}(g)$.
			}}%
	\label{fig:contributions}%
\end{figure}

The QFI of the conditional states $\mathcal{Q}_{s}(g)$ is by the definition the \emph{classical} FI of the optimal measurement $\left\{  \hat{\Pi}%
_{1}^{s},\hat{\Pi}_{2}^{s},...\right\}  $; the probabilities for each outcome to occur are $\left\{  p(1\vert s),p(2\vert
s),...\right\}  ,$ with $p(l\vert s)=$ $\mathrm{Tr}\left[  \rho_{s}\hat{\Pi
}_{l}^{s}\right]  $, therefore:
\begin{equation}
\mathcal{Q}_{s}=\mathcal{F}_{\hat{\Pi}^{s}}=\sum_{l=1}^{L}\frac{1}{p(l\vert s)}\left(  \partial
_{g}p(l\vert s)\right)  ^{2}. \label{25}
\end{equation}
For simplicity we assumed a countable number of outcomes, but everything holds also for POVM labelled by continuous outcomes.
Everything applies also to the distorted state, via the substitution $s \mapsto f$.

The post-selection on the measurement device followed by the optimal measurements on the probe quantum states can be viewed as a POVM performed on the composite state $|\Psi_\mathsf{NLA} \rangle$ : $\ \{\hat{\Pi}_{1}^{s}\otimes\left\vert s\right\rangle \left\langle s\right\vert ,\hat{\Pi}_{2}^{s}\otimes\left\vert s\right\rangle \left\langle s\right\vert,...,\hat{\Pi}_{1}^{f}\otimes\left\vert f\right\rangle \left\langle f\right\vert,\hat{\Pi}_{2}^{f}\otimes\left\vert f\right\rangle \left\langle f\right\vert,...\}$.
The probabilities for each result to appear are $\{p_{s}p(1\vert s),p_{s}p(2\vert s),...,p_{f}p(1\vert f),p_{f}p(2\vert f),...\},$ where $p_{s}$ and
$p_{f}$ as usual represent the probabilities of a successful amplification and a failed one respectively and $p\left(  n\vert i\right)  ,$ $\left(  i=\left\{
s,f\right\}  \right)  $ is the probability that the result of the optimal
measurement $\hat{\Pi}_{n}^{i}$ on the conditional state $\left\vert \psi_{i}\right\rangle $ occurs.

The FI of this probability distribution reads as follows%
\begin{equation}
\mathcal{F}\left[  g\right]  =\sum_{l=1}^{L}\frac{1}{p_{s}p(l\vert s)}\left(
\partial_{g}(p_{s}p(l\vert s))\right)  ^{2}+\sum_{k=1}^{K}\frac{1}{p_{f}p(k\vert
	f)}\left(  \partial_{g}(p_{f}p(k\vert f))\right)  ^{2}, \label{27}%
\end{equation}
which reduces to
\begin{equation}
\mathcal{F}\left[  g\right]  =p_{s} \mathcal{Q}_{s}(g)+p_{f} \mathcal{Q}_{f}(g)+\mathcal{F}_{C},
\label{28}%
\end{equation}
where
\begin{equation}
\mathcal{F}_{C}=\frac{1}{p_{s}}\left(  \partial_{g}p_{s}\right)  ^{2}+\frac
{1}{p_{f}}\left(  \partial_{g}p_{f}\right)  ^{2} \label{29}%
\end{equation}
is the classical FI of the probability distribution $\left\{
p_{s},p_{f}\right\}$.
This quantity is the FI $\mathcal{F}\left[  g\right]$ associated with the post-selection process performed on the composite overall system and it is equal to the QFI of the whole composite system $|\Psi_g \rangle$.
We can see that indeed, our proposed scheme enables to attain the most accurate estimate of the NLA gain.

In~\ref{app:effQFI} we show that the two optimal measurements $\left\{ \Pi^s_k \right\}$ and $\left\{ \Pi^f_k \right\}$ can be chosen to be photon-counting measurements, i.e. $\Pi^{f/s}_k = |k \rangle \langle k |$, independently from the initial state and from the selection outcome.
With some mild assumptions on the initial state also homodyne detection is optimal.
As previously said, these results make our scheme of interest for practical implementations.

In Fig.~\ref{fig:contributions}, we plot the classical FI and QFI that contribute to the
effective QFI as functions of the gain at fixed the threshold and different values of the the input energy $(\overline{n}=1, \overline{n}=2)$.
We note that for a gain greater than 1.3, all the quantities decrease with respect to $g$ and almost vanish for values exceeding $g=4$.
For $g$ varying from 1 to a threshold depending both on the input energy and on the considered probe state, the classical FI arising from the post-selection process is the main contribution to the effective QFI whereas in the remaining parameter region, the amount of information provided by the post-selected amplified states is more substantial.
Finally, we point out a weak contribution coming from the degraded states, particularly in the first region.

Our results indicate that all the steps of the present scheme are important in order to extract the maximum amount of information available on the whole composite state.
In particular, when $g\rightarrow1$ (low amplification regime), the main source of information is the post-selection process itself, i.e: the FI arising from the classical probability
distribution $\{p_{s},p_{f}\}.$

\section{Conclusions}
\label{s:out}

We have addressed the characterisation of non-deterministic 
noiseless linear amplifiers and have compared the performances 
of different estimation strategies aimed at inferring the value 
of the gain, assuming a known threshold. In particular, we 
have analysed minimal implementation of NLA, where the system is
coupled to a two level measuring device via a unitary 
transformation. We have shown that post-selecting only the 
amplified states usually provides a better precision than 
using the unconditional state. On the other hand, the lowest 
quantum Cram\`{e}r-Rao bound is achieved by the extraction 
of all the information contained in the whole composite system.

We have also shown that a feasible sequential measurement scheme 
allows one to access the full information available in the 
overall composite system. In particular, we have found that 
measuring the field or the number of quanta (i.e. homodyne detection
and photon counting for quantum optical implementations) are 
optimal measurements on both conditional states. Assuming to 
have access to an implementation of the NLA, our scheme appears 
to be feasible with current technology, since we only need 
the statistics of the post-selection and a standard homodyne 
measurement on the conditional states.

\appendix
\section{Explicit calculation of the effective QFI}
\label{app:effQFI}

Here we derive the explicit expression of the effective QFI of a generic pure input state.
As we have seen before, the effective QFI is given by the sum of the weighted average of the QFI associated with the states of the mixture plus the FI of the classical probabilities.
Here we show how the the explicit formulas of the involved quantities are derived for a
generic pure state expanded on a Fock basis.

As long as the amplified and distorted states remain pure, their QFI is given by Eq.~(\ref{eq:QFIpure}), i.e. $\mathcal{Q}_{i}(g)=4\left[  \left\langle \partial_{g}\psi_{i}\vert\partial_{g}%
\psi_{i}\right\rangle +\left\langle \partial_{g}\psi_{i}\vert\psi
_{i}\right\rangle ^{2}\right]$ with $i=s,f$.
The overlap of the amplified/distorted state and its derivative with respect
to the gain is found to be null.
The evaluation of the overlap with their
respective derivatives leads to the following expressions
\begin{eqnarray}
&p_{s} \mathcal{Q}_{s}=-\frac{(\partial_{g}p_{s})}{p_{s}}^{2}+4\sum\limits_{n=0}%
^{p-1}(n-p)^{2}g^{2(n-p-1)}\left\vert c_{n}\right\vert ^{2} \\
&p_{f}\mathcal{Q}_{f}=-\frac{(\partial_{g}p_{f})}{p_{f}}^{2}+4\sum\limits_{n=0}%
^{p-1}\frac{(n-p)^{2}g^{4(n-p)-2}}{(1-g^{2(n-p)})}\left\vert c_{n}\right\vert
^{2},
\end{eqnarray}
where the first terms are the same appearing in the classical FI, whereas the second ones are the purely quantum contributions.
As we can see, the effective QFI is reduced to the sum of these purely quantum contributions
\begin{equation}
\mathcal{Q}_\mathsf{eff}(g)=4\sum\limits_{n=0}^{p-1}(n-p)^{2}\left\vert c_{n}\right\vert
^{2}\frac{g^{2(n-p-1)}}{(1-g^{2(n-p)})}.
\end{equation}
We will show that the QFIs of the amplified and distorted state are saturated
by both photon counting and homodyne detection.
The FI associated with that probability distribution reads
\begin{equation}
\mathcal{F}\left[  p_{s}(n\vert g)\right]  =\sum\limits_{n=0}^{\infty}%
\frac{(\partial_{g}\left[  p_{s}(n\vert g)\right]  )^{2}}{\left[  p_{s}(n\vert
	g)\right]  } 
\end{equation}
where $p_s(n\vert g) = |\langle n \vert \psi_s \rangle|^2$.
After summing over all the possible $n$ values, one obtains
\begin{equation}
\mathcal{F}\left[  p_{s}(n\vert g)\right]  =-\left(  \frac{\partial_{g}p_{s}%
}{p_{s}}\right)  ^{2}+\frac{4}{p_{s}}\sum\limits_{n=0}^{p-1}(n-p)^{2}%
g^{2(n-p-1)}\left\vert c_{n}\right\vert ^{2},
\end{equation}
which corresponds exactly to the QFI of the amplified state.

Likewise, the evaluation of the FI associated with a photon counting performed
on the distorted state is carried out starting from
\begin{equation}
p_{f}(n\vert g)= \frac{1}{p_{s}}\left(1-g^{2(n-p)}\right)\left\vert c_{n}\right\vert
^{2},
\end{equation}
from which the following expression of the FI is obtained
\begin{equation}
\mathcal{F}\left[  p_{f}(n\vert g)\right]  =\sum\limits_{n=0}^{p-1}%
\frac{\left(\partial_{g}\left[  p_{f}(n\vert g)\right]  \right)^{2}}{\left[  p_{f}(n\vert
	g)\right]  },
\end{equation}
and where
\begin{equation}
\frac{(\partial_{g}\left[  p_{f}(n\vert g)\right]  )^{2}}{\left[  p_{f}(n\vert
	g)\right]  }=\frac{\left\vert c_{n}\right\vert ^{2}}{p_{f}}\left[
-\frac{\partial_{g}p_{f}}{p_{f}}(1-g^{2(n-p)})-2(n-p)g^{2(n-p)-1}\right]
^{2}.
\end{equation}
Afterwards, summation over $n$ leads to the final FI expression
\begin{equation}
\mathcal{F}\left[  p_{f}(n\vert g)\right]  =-\left(  \frac{\partial_{g}p_{f}%
}{p_{f}}\right)  ^{2}+\frac{4}{p_{f}}\sum\limits_{n=0}^{p-1}\frac
{(n-p)^{2}g^{4(n-p)-2}}{\left(1-g^{2(n-p)}\right)}\left\vert c_{n}\right\vert ^{2}.
\end{equation}
We notice that similarly to the amplified state case, the FI associated with a
photon counting measurement performed on the distorted state saturate its QFI,
thereby showing that a photon counting measurement is optimal regardless on the input data.

Let us also consider homodyne detection with a generic pure input probe state.
The probability distribution associated to a quadrature measurement $\hat{x}=\frac{1}{\sqrt{2}}(\hat{a}+\hat{a}^{\dagger})$ performed on the amplified state reads as follows%
\begin{equation}
p_{s}(x\vert g)=\left\vert \left\langle x\vert\psi_{s}\right\rangle \right\vert^{2}= \left(\frac{2}{\pi} \right)^{1/2}\left(  \sum_{n=0}^{\infty}k_{n}\right)  ^{2},
\end{equation}
where the coefficients are given by 
\begin{equation}
k_{n}=\frac{\exp \left[ -x^{2}\right] }{\sqrt{p_{s}}}
\begin{cases}
\frac{ H_n \left( \sqrt{2}x\right) g^{(n-p)} c_{n} } { 2\sqrt{n!} } \, & \text{if}\; n\leq p \\
\frac{	H_n \left( \sqrt{2} x\right) c_n }{ 2\sqrt{n!}} \, & \text{otherwise}
\end{cases}.
\end{equation}

The evaluation of the FI $\mathcal{F}\left[  p_{s}(x \vert g)\right] = \int_{-\infty}^{\infty} \frac{(\partial_{g}\left[  p_{s}(x\vert g)\right]  )^{2}}{\left[  p_{s}(x\vert g)\right]  }$ is obtained by making of use the properties of the Hermite polynomials $H_n(x)$.
Its expression is found to saturate the QFI under the assumption of real coefficients of the input generic pure state\footnote{Clearly, if the probe state is rotated with the unitary operator $e^{i \theta \hat{n}}$, the bound is still saturated, but the optimal homodyne measurement is rotated as well (i.e. one has to measure a different quadrature).}.

As for the amplified state, the expression of the FI associated with a homodyne detection applied on the distorted state is derived starting from the following probability distribution
\begin{equation}
p_{f}(x\vert g)= \left(\frac{2}{\pi}\right)^{1/2}\frac{\exp\left[  -2x^{2}\right]
}{p_{s}}\left(  \sum_{n=0}^{\infty}\frac{H_{n}\left(  \sqrt{2}x\right)
	\sqrt{1-g^{2(n-p)}}c_{n}}{2\sqrt{n!}}\right)  ^{2},
\end{equation}
and found to saturate the QFI.

Similarly to the photon counting measurement, the evaluation of the FIs
associated to the probability distributions resulting from a quadrature
measurement performed on both conditional states reveal that this
latter is optimal with the requirement of the coefficients of the input pure
state to be real.

\section{Estimation with a generic state of the meter qubit}
\label{app:genericMeter}
Here we compute the QFI of the state obtained by acting with the global unitary~\eqref{16} on the tensor product of the input state and an arbitrary state of the meter qubit.
Let us write again the global unitary~\eqref{16}, derived in~\cite{McMahon}
\begin{equation}
\hat{U}_g = \hat{E}_s^p \otimes | s \rangle \langle f | + \hat{E}_f^p \otimes | s \rangle \langle f | - \hat{E}_s^p \otimes | f \rangle \langle s | + \hat{E}_f^p \otimes | s \rangle \langle s | \,,
\end{equation}
we don't need the explicit form of the two Kraus operators~\eqref{6} and~\eqref{7}, but we need to remember that they are Hermitian and commuting, since they are both diagonal in Fock basis.

The state that we are considering for this analysis is the following one
\begin{equation}
\begin{split}
| \Psi_g  \rangle &= \hat{U}_g \left[ | \psi \rangle \otimes \left( \alpha | s \rangle + \beta | f \rangle \right) \right] = \\
&= \left( \beta \hat{E}_s^p | \psi \rangle + \alpha \hat{E}_f^p | \psi \rangle  \right)\otimes  |s \rangle  + \left( \beta \hat{E}_f^p | \psi \rangle - \alpha \hat{E}_s^p | \psi \rangle  \right)  \otimes  | f \rangle \\
& = | \tilde{\phi}_s \rangle \otimes |s \rangle + | \tilde{\phi}_f \rangle \otimes |f \rangle
\end{split},
\end{equation}
where $|\alpha|^2 + |\beta|^2 = 1 $; for $\beta=1$ and $\alpha=0$ we get the state $|\Psi_\mathsf{NLA} \rangle$ considered in the main text, see Eq.~\eqref{eq:FullPureNLA}.
We have also introduced two unnormalized states: $| \tilde{\phi}_s \rangle = \beta \hat{E}_s^p | \psi \rangle + \alpha \hat{E}_f^p | \psi \rangle $ and $| \tilde{\phi}_f \rangle = \beta \hat{E}_f^p | \psi \rangle - \alpha \hat{E}_s^p | \psi \rangle$.
To compute the QFI we need the derivative of this pure state w.r.t. to $g$:
\begin{equation}
\begin{split}
| \partial_g \Psi_g  \rangle &= \partial_g \hat{U}_g \left[ | \psi \rangle \otimes \left( \alpha | s \rangle + \beta | f \rangle \right) \right] = \\
&= \left( \beta \partial_g \hat{E}_s^p | \psi \rangle + \alpha \partial_g  \hat{E}_f^p | \psi \rangle  \right)\otimes  |s \rangle  + \left( \beta \partial_g  \hat{E}_f^p | \psi \rangle - \alpha \partial_g  \hat{E}_s^p | \psi \rangle  \right)  \otimes  | f \rangle \\
& = | \partial_g \tilde{\phi}_s \rangle \otimes |s \rangle + | \partial_g \tilde{\phi}_f \rangle \otimes |f \rangle
\end{split},
\end{equation}
where the derivative of the Kraus operators $\partial_g \hat{E}_s^p$ and $\partial_g \hat{E}_f^p$ remain diagonal in Fock basis and thus they satisfy the relationship
\begin{equation}
\hat{E}_s^p \partial_g \hat{E}_s^p + \hat{E}_f^p \partial_g \hat{E}_f^p  = 0, \label{eq:derKraus}
\end{equation}
which comes from the taking the derivative of the equality $ \hat{E}_{s}^{p}{}^\dag \hat{E}_{s}^{p} + \hat{E}_{f}^{p}{}^\dag \hat{E}_{f}^{p} = \mathbbm{1}$.

The first term of the QFI is found to be
\begin{align}
\langle \partial_g \Psi_g | \partial_g \Psi_g \rangle & = \langle \partial_g \tilde{\phi}_s | \partial_g \tilde{\phi}_s \rangle + \langle \partial_g \tilde{\phi}_f | \partial_g \tilde{\phi}_f \rangle =\\
&=\left( |\alpha|^2 + |\beta|^2 \right) \left[ \left\langle \psi \left| \left( \partial_g \hat{E}_s^p \right)^2 \right| \psi \right\rangle + \left\langle \psi \left| \left( \partial_g \hat{E}_f^p \right)^2 \right| \psi \right\rangle\right] =\\
& = \left\langle \psi \left| \left( \partial_g \hat{E}_s^p \right)^2 \right| \psi \right\rangle + \left\langle \psi \left| \left( \partial_g \hat{E}_f^p \right)^2 \right| \psi \right\rangle,
\end{align}
it is straightforward to verify that these two terms are equal to the second terms in Eqs.~(\ref{21},\ref{22}) and thus $\langle \partial_g \Psi_g | \partial_g \Psi_g \rangle = \mathcal{Q}_\mathsf{eff} (g)$ regardless of the state of the meter qubit.

On the other hand, the second term of the QFI is found to be
\begin{align}
\langle \Psi_g | \partial_g \Psi_g \rangle &= \langle \tilde{\phi}_s | \partial_g \tilde{\phi}_s \rangle + \langle \tilde{\phi}_f | \partial_g \tilde{\phi}_f \rangle \\ 
&= 2 i \mathrm{Im} \left[ \alpha \beta^* \right] \left( \left\langle \psi \left| \hat{E}_s^p \left( \partial_g \hat{E}_f^p \right) \right| \psi \right\rangle - \left\langle \psi \left| \hat{E}_f^p \left( \partial_g \hat{E}_s^p \right) \right| \psi \right\rangle \right) \,, \label{eq:genMeter2ndterm}
\end{align}
where some terms disappear because of~\eqref{eq:derKraus}.
The formula for the QFI is $\mathcal{Q}\left( |\Psi_g \rangle \right) =\langle \partial_g \Psi_g | \partial_g \Psi_g \rangle - \left| \langle \Psi_g | \partial_g \Psi_g \rangle \right|^2$ and when the term~\eqref{eq:genMeter2ndterm} is not zero it always decreases the magnitude of the QFI; therefore we have the inequality stated in the main text:
\begin{equation}
\mathcal{Q}\left( |\Psi_g \rangle \right) \leq \mathcal{Q}\left( |\Psi_\mathsf{NLA} \rangle \right) = \mathcal{Q}_\mathsf{eff}(g).
\end{equation}
\ack
This work has been supported by SERB through project VJR/2017/000011. 
The authors are grateful to Matteo Bina and Luigi Seveso for useful discussions.
MGAP is member of GNFM-INdAM.
FA acknowledges support from the UK National Quantum Technologies Programm (EP/M013243/1)
\section*{References}
\bibliographystyle{iopart-num}
\bibliography{Ref}
\end{document}